\documentclass[prd, preprint, aps]{revtex4}

\usepackage{graphicx}
\usepackage{color}
\usepackage{amsfonts}

\renewcommand{\c}{\cos}
\renewcommand{\th}{\theta}
\renewcommand{\b}{\bar}
\renewcommand{\t}{\bar}
\newcommand{\f}{\frac}
\newcommand{\p}{\perp}
\newcommand{\bt}{\beta}
\newcommand{\s}{\sin}
\newcommand{\ph}{\phi}
\newcommand{\g}{\gamma}

\newcommand{\be}{\begin{equation}}
\newcommand{\ee}{\end{equation}}
\newcommand{\ba}{\begin{eqnarray}}
\newcommand{\ea}{\end{eqnarray}}
\newcommand{\beq}{\begin{eqnarray}}
\newcommand{\eeq}{\end{eqnarray} }
\begin{document}

\title{Area Invariance of Apparent Horizons under Arbitrary Lorentz Boosts}

\author{Sarp Akcay}
\author{Richard A. Matzner}
\author{Vishnu Natchu}

\affiliation{ University of Texas at Austin }
            \email{akcays@physics.utexas.edu}

            \email{matzner2@physics.utexas.edu}

           \email{vishnu@physics.utexas.edu}


\begin{abstract}
It is a well known analytic result in general relativity that the
2-dimensional area of the apparent horizon of a black hole remains
invariant regardless of the motion of the observer, and in fact is
independent of the $ t=constant $ slice, which can be quite
arbitrary in general relativity. Nonetheless the explicit
computation of horizon area is often substantially more difficult
in some frames (complicated by the coordinate form of the metric),
than in other frames. Here we give an explicit demonstration for
very restricted metric forms of (Schwarzschild and Kerr) vacuum
black holes. In the Kerr-Schild coordinate expression for these
spacetimes they have an explicit Lorentz-invariant form. We
consider {\it boosted} versions with the black hole moving through
the coordinate system. Since these are stationary black hole
spacetimes, the apparent horizons are two dimensional cross
sections of their event horizons, so we compute the areas of
apparent horizons in the boosted space with (boosted) $ t =
constant $, and obtain the same result as in the unboosted case.
Note that while the invariance of area is generic, we deal only
with black holes in the Kerr-Schild form, and consider only one
particularly simple change of slicing which amounts to a boost.
Even with these restrictions we find that the results illuminate
the physics of the horizon as a null surface and provide a useful
pedagogical tool. As far as we can determine, this is the first
explicit calculation of this type demonstrating the area
invariance of horizons. Further, these calculations are directly
relevant to transformations that arise in computational
representation of moving black holes. We present an application of
this result to initial data for boosted black holes.

\end{abstract}
\maketitle
\section{Introduction}\label{sec:intro}
\label{intro} Apparent horizons (AH) were first introduced by
Penrose and Hawking \cite{PH}, \cite{HE}. An apparent horizon is
defined as the outermost marginally trapped surface on a given
(partial) Cauchy slice. It is a topologically spherical
2-dimensional surface on which the expansion of the outgoing null
rays orthogonal to the surface is zero \cite{BBGA}. Thus, it is a
surface where gravity is so strong that putative outgoing null
rays can only ``hover" against the gravitational force. Unlike
event horizons, which are globally defined as the boundary in
spacetime between null geodesics that escape to infinity, and
those that fall into the singularity, apparent horizons are local
objects, computable at one instant of time, hence much more
accessible in numerical simulations. The locations of event and
apparent horizons coincide only in stationary spacetimes. In the
stationary Kerr spacetime in Boyer-Lindquist coordinates, the
horizon is located at radial coordinate $ r = r_+ \equiv M +
\sqrt{M^2 - a^2} $, where $ M $ is the mass of the black hole and
$ a $ is the spin parameter for the Kerr black hole given by $ a
\equiv J/ M $, with $ J $ being the angular momentum of the black
hole; for $a=0$ we have the static Schwarzschild black hole. Here
and henceforth, we use Newton's constant $ G = 1$, and speed of
light $c=1$.

In this paper we consider only Kerr spacetimes in Kerr-Schild (KS)
coordinates as given in Eq. (\ref{eq:KSmetric}) below. This form
of the metric contains a ``natural" Minkowski background, and
hence a natural definition of a Lorentz boost \cite{KS}. It is
found (cf. \cite{Hu}, \cite{Ma}, \cite{HCM}) that the apparent
horizon of a black hole will appear distorted in these coordinates
when boosted; the longitudinal coordinate direction undergoes a
Lorentz contraction. However, this is an effect only in
coordinates; the point of this paper is an explicit calculation to
show that the area of the apparent horizon 2-surface, recomputed
in the spatial frame of the boosted observer, remains unchanged,
that is: $ Area = 4 \pi\left(r_+^2+ a^2 \right) $ for the Kerr
case and $ Area = 16 \pi M^2 $ for the Schwarzschild black hole.
This result is of course necessary on general principles.
\par The invariance of the area depends on
the observation that the event horizon of a stationary black hole
is a null 3-dimensional submanifold of the spacetime with
vanishing expansion. And null surfaces naturally remain null under
Lorentz transformations. In fact, the area of any 2-dimensional
cross section of the horizon remains invariant under any
redefinition of the 3-space $ t = constant $ (that is legitimately
spacelike). Two cross sections of the event horizon that differ by
a redefinition of $ t =constant $ slice can be put in a pointwise
1-to-1 correspondence along the null generators of the horizon.
These null offsets do not contribute to the area which is
transverse to the null generators. We give a quick derivation of
the Schwarzschild situation and then present the most general
calculation for these spacetimes, namely, the Kerr black hole
boosted along an arbitrary direction.
\par The Kerr vacuum solution to Einstein's equation can be written
in a special form called the Kerr-Schild form of the metric. This
form is, in general (\cite{Po}, \cite{MTW}, \cite{Ho}, \cite{Ch}),
\be g_{\mu \nu} = \eta_{\mu \nu} + 2 H l_{\mu} l_{\nu}
\label{eq:KSmetric} \ee
where $ H $ is a function of spacetime coordinates, $ \eta_{\mu
\nu} $ is the Minkowski metric of flat spacetime and $ l^{\mu} $
is a null vector with respect to both $ g_{\mu \nu} $ and $
\eta_{\mu \nu} $. Clearly, this is a special form, and the metric
of a general spacetime cannot  be put in this form. But the Kerr
vacuum black hole can be so written. Under a Lorentz boost (a
coordinate transformation with the form of a Lorentz
transformation on the $t,x,y,z$ coordinates describing the flat
space with metric $\eta_{\mu \nu } dx^{\mu}dx^{\nu}) $, the
Kerr-Schild metric will preserve the general form that it has in
Eq. (\ref{eq:KSmetric}). We will place overbars on coordinates in
the unboosted frame. In section \ref{sec:KerrBH}, we will show the
area invariance for a boosted Kerr black hole by performing a
coordinate transformation to facilitate boosting the spacetime,
followed by another coordinate transformation that simplifies
extracting the 2-dimensional metric by restricting to the horizon.
With the 2-dimensional metric we straightforwardly compute the
horizon area.

The special case of the nonspinning Schwarzschild ({\i.e.}
spherical) black hole provides an illuminating guide to the
features of the full Kerr case. Eq. (\ref{eq:KSmetric}) for this
case is
\be g_{\mu \nu}d\b{x}^{\mu}d\b{x}^{\nu} = -d\b{t}^2 + d\b{x}^2 +
d\b{y}^2 + d\b{z}^2 + \f{2M}{\b{r}}  \left(d\b{t} + d\b{x} +
d\b{y} + d\b{z} \right)^2 \label{eq:KsSCHmetric} \ee
which, in cylindrical coordinates $ (\b{r}_{||}, \b{r}_\perp,
\b{\phi}_{cyl} ) $ can be written as
\be g_{\mu \nu}d\b{x}^{\mu}d\b{x}^{\nu} = -d\bar{t}^2 +d \bar
r_{||}^2 +d\bar x_\perp^2 + d\bar y_\perp^2 + \frac{2M}{\bar
r}\left(d\bar{t} + \f{\b{r}_{||}}{\b{r}} d \bar r_{||} +
\f{\b{x}_\perp}{\b{r}} d\bar x_\perp +\f{\b{y}_\perp}{\b{r}} d\bar
y_\perp\right)^2 . \label{eq:Schwarzmetric} \ee
where $\b{x}_\perp = \b{r}_\perp cos\: \b{\phi}_{cyl}$ and
$\b{y}_\perp = \b{r}_\perp sin\: \b{\phi}_{cyl}$.  The coordinate
system $(\b{r}_{||}, \b{r}_\perp, \b{\phi}_{cyl})$ aligns $
\b{r}_{||} $ with the axis of the cylinder parallel to the boost
direction $ \vec{\beta} $; $ \b{r}_\p, \b{\ph}_{cyl} $ are the
polar coordinates of the circular plane orthogonal to the axis of
the cylinder (see Fig. (\ref{fig:fig1})). Note that $ \b{x}^2 +
\b{y}^2 + \b{z}^2 = \b{r}_{||}^2 + \b{r}_\p^2 $.
\par The boosted
(unbarred) coordinates are related to the unboosted frame by
\ba \b{t} & = & \gamma (t - \bt \; r_{||} ) \nonumber \\
\b{r}_{||} & = & \g ( r_{||} - \bt \; t) \nonumber \\ \b{r}_\p & =
& r_\p \; , \quad \b{\ph}_{cyl} = \ph_{cyl}.
\label{eq:LorentzTrans} \ea
The boost parameter is $\beta\equiv v/ c$, and $\gamma =
(1-\beta^2)^{-1/2}$; both are defined as usual in the background
Minkowski spacetime. The apparent horizon is defined in a given
3-space ($t=constant$) and the horizon area will be independent of
$t$, so we take $t=0$. The $t=0$ (boosted) 3-metric is
\be ds^2{\mid}_{t = 0} = d r_{||}^2  +d x_\perp^2 + dy_\perp^2 +
\frac{2M}{\bar r^3}(-\bar r\gamma \beta dr_{||} +\bar r_{||} d\bar
r_{||}  +\bar x_\perp d\bar x_\perp + \bar y_\perp d\bar
y_\perp)^2\: .  \ee
We have strategically kept some terms expressed using unboosted
(barred) forms. They can be straightforwardly substituted using
Eq. (\ref{eq:LorentzTrans}). In this form, however, we can easily
restrict the metric to the horizon surface, since
\be \bar rd\bar r= \bar r_{||} d\bar r_{||} +\bar x_\perp d\bar
x_\perp + \bar y_\perp d\bar y_\perp = 0 \label{eq:rdr} \ee
on the horizon where $\bar r$ is a constant ($= 2M$). Thus on the
horizon:
\begin{eqnarray}
ds^2{\mid}_{t = 0,\: \b{r}=2M} &=& \left[\gamma^{-2} d\bar
r_{||}^2 +d \bar x_\perp^2 + d \bar y_\perp^2 +
\frac{2M}{\bar r}(\beta^2 d\bar r_{||}^2)\right]_{\b{r}=2M}\nonumber \\
&= & (d\bar r_{||}^2  +d \bar x_\perp^2 + d \bar
y_\perp^2)|_{\b{r}=2M}\: . \label{eq:ds2_horizon}
\end{eqnarray}
In Cartesian $ (\b{x}, \b{y}, \b{z}) $ coordinates this would look
like
\be ds^2{\mid}_{t = 0,\: \b{r}=2M} = \left(d\b{x}^2 + d\b{y}^2 +
d\b{z}^2 \right)|_{\b{r}=2M}\: .\label{eq:ds2_horizon2} \ee
This can be put in a more familiar form using spherical
coordinates $ (\b{r}, \b{\theta}, \b{\phi}) $ which now gives
\be ds^2{\mid}_{t = 0,\: \b{r}=2M} = (2M)^2 \left(d\b{\theta}^2 +
\sin^2 \b{\theta} d\b{\phi}^2 \right)\: . \ee
Thus the area of the horizon is $4 \pi (2M)^2$ as expected.
Importantly, note that Eq. (\ref{eq:ds2_horizon}) describes the
boosted apparent horizon; the simple form (Eq.
(\ref{eq:ds2_horizon2})) that allows immediate evaluation of the
surface area is the expression of this area in terms of
coordinates appropriate first of all to the unboosted frame. On
the horizon the contribution from the time transformation exactly
cancels the Lorentz contraction of $ \b{r}_{||} $.

\section{Numerical Results}
Before looking at the horizon of a boosted spinning black hole, we
demonstrate some numerical applications of these concepts,
concentrating in this section on only nonspinning black holes.
Recent breakthroughs in numerical relativity (\cite{Pr},
\cite{BCCKM}, \cite{CLMZ}, \cite{GHSBH}, \cite{CLZM},
\cite{BGHHS}) have enabled the community to investigate various
physical scenarios involving interacting black holes. There are
many different approaches to numerically evolving the physical
system. The use of a particular structure, {\it puncture initial
data} (\cite{BB}) has become ubiquitous for numerical codes.
Puncture initial data are conformally flat. Solution of the
constraint equations (elliptic equations describing a nonlinear
generalization of Newtonian gravity) produces a mathematically
correct configuration. But if boosted, the puncture is not
physically relaxed, so when the solved (mathematcally correct)
data are evolved, the black hole emits short wavelength
gravitational radiation. Some of this spurious radiation
propagates out to infinity and some falls onto the black hole,
increasing the horizon mass.
\par One can instead use {\it superposed Kerr-Schild} (\cite{MaIII}) initial
data. This takes the Kerr-Schild metric for a single black hole
and creates a background metric for two black holes by adding a
second `mass term' to the flat background:
\be g_{\mu\nu} = \eta_{\mu\nu} + H_1 l^{(1)}_{\mu} l^{(1)}_{\nu} +
H_2 l^{(2)}_{\mu} l^{(2)}_{\nu}. \label{eq:SuperKS} \ee
Here $ H_1, H_2 $ are scalar functions that depend on coordinates
from the centers of each black hole as well as the black holes'
masses and spins. They are identical in form to single black hole
terms centered at the locations of the two holes (cf. Eqs. (1),
(2), and also Eq. (11) below; there is also a prescription for
superposing the momentum associated with this combination, in the
initial data). Although Kerr-Schild initial data exactly solve
Einstein's equation for a single boosted black hole and thus
satisfy the constraint equations, this is not the case for
superposed Kerr-Schild, which is only an educated guess. However,
by starting out with this initial guess as a conformal background
metric (in the same sense that puncture data has a flat conformal
background), one can solve the constraint equations, so
Kerr-Schild data can be adjusted to become proper initial data.
The solution of the elliptic initial data equations modifies the
configuration to be an exact (modulo numerical error) description
of a gravitational configuration. In practice, unless the black
holes are very close together, the correction for superposed
Kerr-Schild data is small; less than one percent.
\par The code being developed at University of Texas Austin is
called \emph{openGR} \cite{MNW}. Among the suite of programs
comprising {\it openGR}, there is a finite element initial data
code, which can produce either puncture or superposed Kerr-Schild
initial data. The evolution code treats the dynamics of binary
black hole systems and the extraction of gravitational waves from
the merger of the black holes. The code is a fourth order accurate
adaptive mesh refinement code with sixth order interpolation
between coordinate patches.

The total mass/energy of the spacetime is given by the ADM mass $
M_{ADM} $ (\cite{ADM}) computed at spatial infinity (numerically,
``near" the grid boundary). The ADM mass corresponds to the
apparent Newtonian mass measured at large distances from the
sources, measured for instance by observing the period of distant
satellites around the central mass. Suppose the individual black
hole masses are given by Kerr-Schild mass parameters $ m_1 $ and $
m_2 $. Then the {\it background} gives an ADM mass $ M_{ADM~bkgd}
= m_1 + m_2 $. As noted, solving the constraint equation changes
the superposed Kerr-Schild data slightly, so the solved ADM mass
closely approximates  $ M_{ADM} \approx m_1 + m_2 $, though it
does have some dependence on the parameters of the data,
particularly on the separation of the black holes.

Of interest in the design of data is the binding energy $ E_b $ of
the configuration. We can compute this as the measured ADM mass
minus the intrinsic mass of the constituent black holes. The
difficulty lies in defining an {\it } intrinsic black hole mass.
We choose the horizon mass. (For nonspinning black holes, we have
$M_H= (A_H/16 \pi)^{1/2}$, where $A_H$ is the area of the apparent
horizon; $openGR$ includes an apparent horizon finder.)
Classically the area of the horizon can increase, but we also know
that the horizon area is an adiabatic invariant; it is only
slightly affected by slow motions. ``Slow" means slow compared to
the normal frequencies of oscillation of the hole, which are high
frequency; the lowest frequency is on the order $f \sim (20
M_H)^{-1}$, and most frequencies in binary evolution are lower
than this frequency. Hence we are confident that the apparent
horizon provides an (almost) constant intrinsic mass. Binding is
indicated by $ E_b \equiv M_{ADM} - ( M_{H1} +M_{H2}) < 0$. If the
data describe black holes in motion, then the kinetic energy also
contributes (positively) to the total energy. For a boosted black
hole the ADM mass acquires a factor $\gamma$: $M_{ADM~0}
\rightarrow \gamma M_{ADM~0}$ where $M_{ADM~0}$ is $m$, the metric
mass parameter in the single hole case. Thus we expect that for a
given boost parameter $\gamma$, the binding energy may be negative
(i.e. bound) if the holes are close together, but positive
(unbound) if the data are set with the black holes far apart.
Furthermore, for the nonlinear small separation limit (and/or for
significant $\gamma$) cases, Newtonian arguments become obscure
because of the change in metric due to the presence of the second
hole, and due to coordinate ambiguities.

We construct an equal mass binary black hole system (nonspinning
Schwarzschild black holes) with initial coordinate separation $r$.
The configuration is axisymmetric; the black holes are boosted
toward or away from each other with Lorentz boost velocity $ \beta
\equiv v/ c $ (or instantaneously at rest with $\beta = 0$); the
boosts are equal but opposite in the computational frame. The
axisymmetry allows extremely high resolution computational
simulation. The code is a finite element code, with an adaptive
resolution of $1/100~ M_{ADM}$ near the holes and $1 \ M_{ADM}$ at
the outer boundary. The computational domain is a sphere of radius
$256~ M_{ADM}$.

\begin{figure*}
\includegraphics[height=10cm]{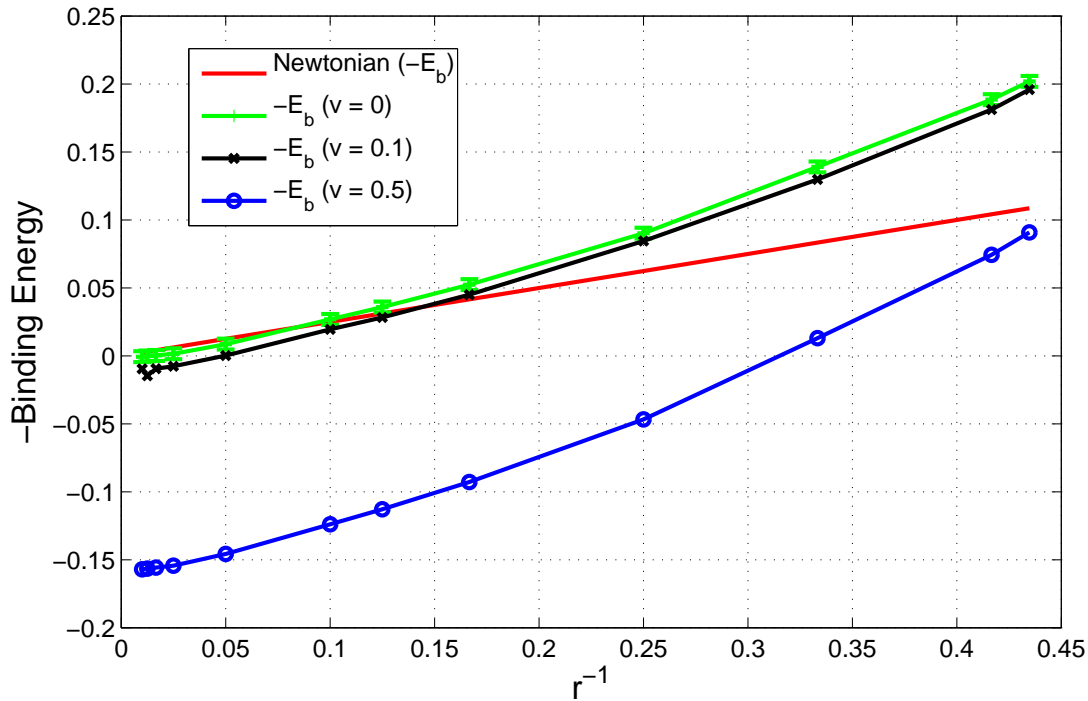}
\caption{Negative of the binding Energy $-E_b$ versus the inverse
coordinate separation $1/ r$ for the cases with boosts speed
$\beta=0, ~0.1,~0.5$ represented by the green, black and blue
curves, respectively. The red line is the Newtonian binding energy
which scales as $1/ r$. Ideally, it should be tangent to the
$\beta=0$ curve (green) at large $r$ ($1/r\rightarrow 0$) but here
it is slightly shifted due to numerical errors. As can be seen in
the figure, the binding energy matches the Newtonian limit very
well for large separations ($1/ r\rightarrow 0$), it grows faster
than $1/ r$ as the black holes are closer ($1/
r\rightarrow\infty$). This is due to changes in horizon masses
because of the distortions induced by the black holes on each
other. It ($-E_b$) also becomes more negative for large boosts
reflecting the unbound nature of distant rapidly moving black
holes. The kinetic energy of the black holes overwhelms the
negative potential energy.}\label{fig:bigVish1}
\end{figure*}

We plot the the negative of binding energy $-E_b$ of the binary in
figure \ref{fig:bigVish1}. Here, we define the binding energy to
be $E_b \equiv M_{ADM} - 2 M_H$. Since the configuration is
axisymmetric, the black holes have the same horizon mass $M_H$,
hence the factor of 2. We display our results in units of the
total parameter ADM mass which is normalized to equal 1
($M_{ADM~bkgd}=m_1+m_2 = 0.5+0.5=1$). 
Here $r$ is the coordinate separation between the two black holes
also given in units of $M_{ADM}$ (e.g. $r=10$ translates to $r=
10\:G\:M_{ADM}/c^2)$. The binding energy scales as $1/ r$ at the
Newtonian limit; this Newtonian limit is plotted as a red straight
line in Fig. \ref{fig:bigVish1}. Bonning et al. \cite{Bonningetal}
had analytically predicted this Newtonian limit (see that paper
for details). Previous computational work by Hawley et al.
\cite{Hawleyetal} failed to show the Newtonian limit, because of
insufficient domain size to eliminate outer boundary effects. We
clearly see that for every rest configuration the binding energy
for large separations agrees with the Newtonian prediction
(\cite{Bonningetal}), but there is a deviation to stronger binding
for closer coordinate separation. We will have more to say about
this in a future paper. The cause for this will be discussed below
as it related to the distortion of black holes' horizons near each
other. One can in principle use expressions from post-Newtonian
theory to give the next order correction to $ E_b $. These terms
scale as $ (\mathrm{(Mass)}/ r)^2$. We have begun studying these
higher order corrections.

\begin{figure*}
\includegraphics[height=10cm]{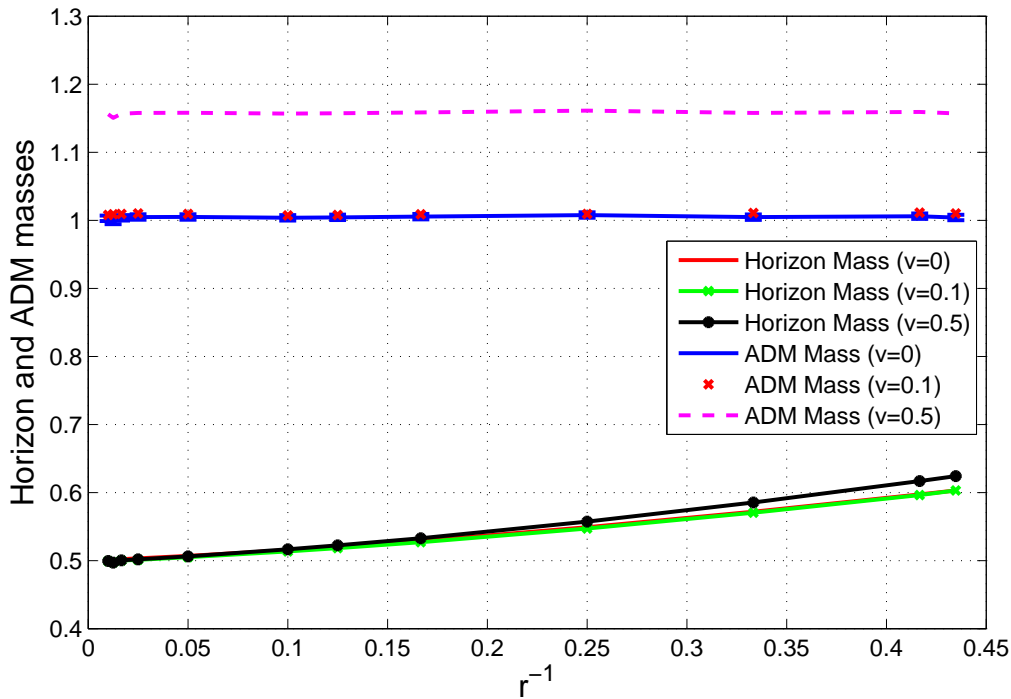}
\caption{Horizon mass $M_H$ and ADM mass $M_{ADM}$ versus inverse
distance 1/r for boost speeds of $\beta= 0, 0.1, 0.5$. $M_{ADM}$
(the upper, approximately parallel curves) is given by the the
pink dashed line (v=0.5) and by the closely overlapping blue line
(v=0) and the red tick marks (``$\times$"). As expected, the ADM
mass remains constant regardless of the separation r but varies as
$\gamma \ M_{ADM \ 0}$ for varying boost speeds $\beta$. $M_H$
(the lower curves) is represented by closely overlapping  red and
green curves for the v=0 and v= 0.1 cases, and by the higher black
curve for v= 0.5. Note that the horizon mass grows larger as the
black holes are nearer i.e. as $1/r \rightarrow \infty$. The
horizon mass is invariant under boosts. For $r \ge 10$ (i.e. $1/r
\le 0.1$) the horizon mass curves for different boosts overlap
perfectly. Apparently because of the nonlinear interaction of the
black hole geometries in the full solution, for larger boosts and
for small separations the horizon mass does increase
slightly.}\label{fig:bigVish2}
\end{figure*}

\begin{figure*}
\includegraphics[height=8cm, angle=330]{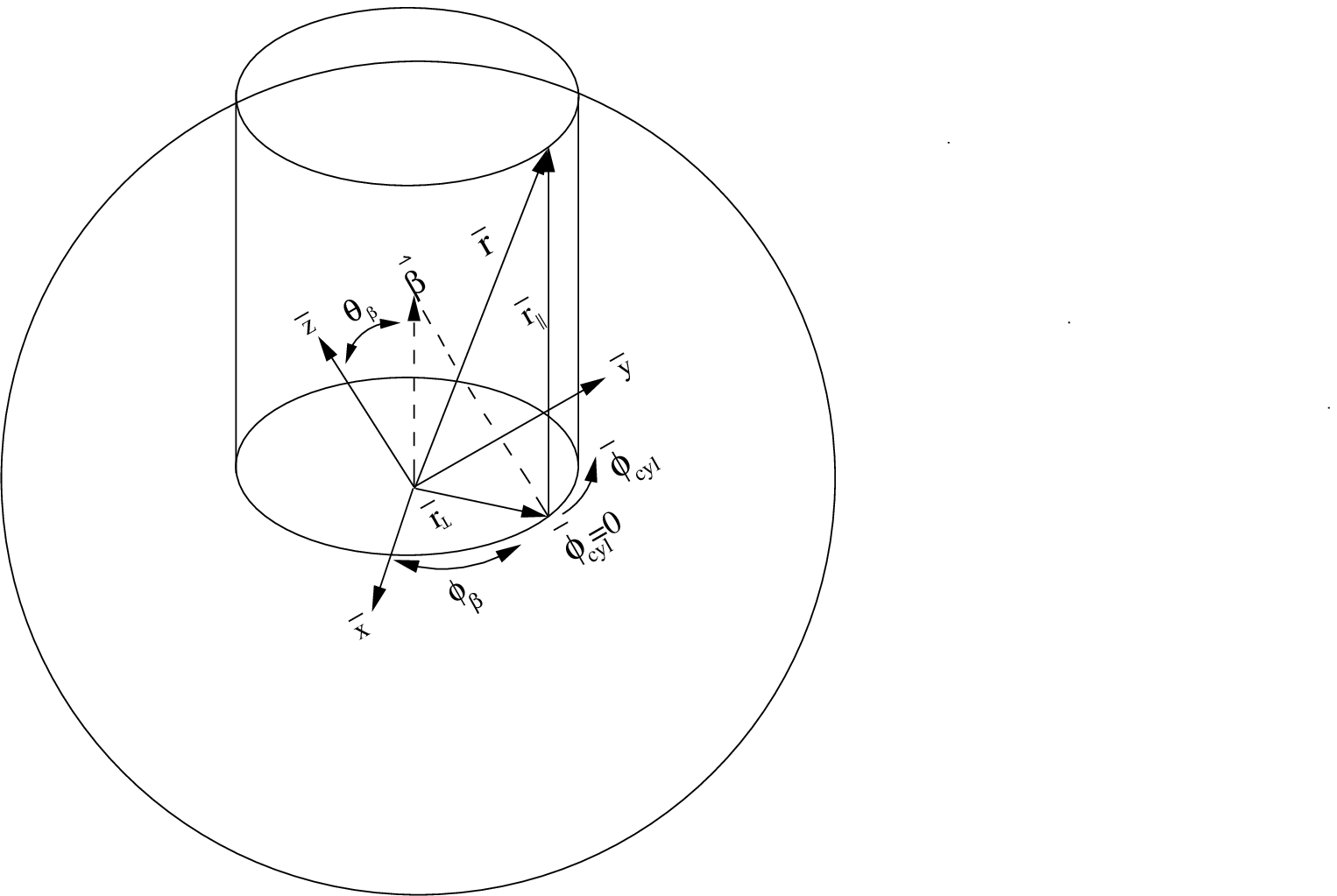}
\caption{The tilted cylindrical coordinates $ (\b{r}_{||},
\b{r}_\perp, \bar{\phi}_{cyl}) $ along with the radial coordinate
$ r $ and Kerr-Schild Cartesian coordinates $ (\b{x}, \b{y},
\b{z})$. The vector $ \mathbf{\beta} $ points along the boost
direction, which is parallel to the symmetry axis of the
cylinder.} \label{fig:fig1}
\end{figure*}

It is of interest to understand {\it how} the binding energy is
achieved in the initial data. Fig. \ref{fig:bigVish2} is a plot of
$M_{ADM}$ and horizon mass $M_H$ versus $ 1/ r $ for boosts of
$\beta = 0, ~0.1, ~0.5$ represented by the red, green and black
curves for $M_H$ , respectively. For the $M_{ADM} $ versus $1/ r$
plot, we use a blue solid line, red ``$\times$'' marks and pink
dashed line for $\beta = 0, ~0.1, ~0.5$, respectively. Note the
confirmation of the analytical expectation above that the ADM mass
is essentially constant for the binary pair regardless of the
coordinate distance between them. However, although we construct
all data with the same parameter values $m$, we see different
constant ADM masses for different $|\beta|$ (motion with the same
$|\beta|$ together or apart yields the same ADM mass, constant
across the possible separations). This is because the ADM mass
scales as $\gamma \ M_{ADM\:0}$ for a boosted black hole. Thus,
for example, the ratio of ADM masses between the pink dashed line
($\beta=0.5$) and the blue line ($\beta=0$) in Fig.
\ref{fig:bigVish2} should be $\left(1-0.5^2\right)^{-1/2} =
1.154$. This is easily seen in Fig. \ref{fig:bigVish2}. We
estimate the numerical error of about one percent in this quantity
by looking at the ADM mass for the $\beta=0$ case (blue line)
which, in principle, should give $M_{ADM}=1$ but actually is
located slightly higher at $M_{ADM} = 1.01$.

Though $M_{ADM}$ stays almost constant for differing separation,
the binding becomes stronger for smaller separation, even in the
Newtonian limit, of course. As described in \cite{Bonningetal},
when the parameter $m$ is held constant for each hole, the {\it
horizon} area of the constituent black holes increases with
decreasing separation. The modification of the geometry by the
{\it other} black hole modifies the horizon area so that it is no
longer the $16 \pi m^2$ which would be computed for an isolated
hole, but $16 \pi M_H^2$ with $M_H \ne m$. If we imagine the
initial data constructed by adiabatically moving the holes from
infinite separation, it would be this mass $M_H$ which is
adiabatically invariant. This was predicted analytically by
\cite{Bonningetal} for nonspinning, instantaneously nonmoving
black holes; it was predicted qualitatively for moving black
holes.

The horizon mass is expected to remain invariant under boosts, and
in single boosted black holes this is what we observe. But for
fully solved data -- the result of solving a nonlinear elliptic
system, Fig. \ref{fig:bigVish2} shows that the horizon mass in the
$\beta=0.5$ case is somewhat above that of the $\beta=0$ one for
close separations (`close' meaning black hole separations less
than $r=10$). Indeed, for $r>10$ (i.e. $1/ r < 0.1$), the overlap
of the red, green and black curves is perfect to within less than
one percent error. This is an interesting result depending on both
the boost and the separation. The \emph{growth} of the horizon
area for large boosts is an effect due to the proximity of the two
black holes. For sufficiently large separations, the boost does
not change the horizon mass, hence the horizon area.

\section{Boosted Kerr Black Hole}\label{sec:KerrBH}
We return to the analytic study of black hole horizons, now
including spin. Strong astrophysical evidence supports the
existence of spinning (Kerr) black holes (\cite{Ge},
\cite{Na},\cite{BFDL}); manipulating description of this spacetime
is a frequent task in computational astrophysics. The angular
momentum of the spinning black hole automatically selects a
preferred direction and the Kerr hole is axially symmetric around
the spin axis. Written in Kerr-Schild coordinates the Kerr
spacetime formally admits a boost.

We will begin with the unboosted Kerr metric written in standard
Kerr-Schild coordinates. We will then rewrite the metric in
cylindrical coordinates where the symmetry axis of the cylinder
points toward the boost direction. (We transform to cylindrical
coordinates only to facilitate the boosting of the spacetime.)
Once the spacetime is boosted, we will look at the spatial
3-metric on a (boosted) $ t = constant $ hypersurface. Since we
are ultimately interested in the 2-metric we will perform one
final coordinate transformation from cylindrical to spheroidal
coordinates and consider $\b{r} = r_+$ (the expected horizon
location). Once we have our 2-metric, we will compute the area of
the apparent horizon and show that it indeed equals the unboosted,
stationary value, which is $ Area = 4 \pi \left(r_+^2+ a^2 \right)
$.

\par
The Kerr spacetime in Kerr-Schild coordinates is (\cite{Po},
\cite{MTW}, \cite{Ho},\cite{Ch}):
\ba ds^2 & = & -d\b{t}^2 + d\b{x}^2 + d\b{y}^2 + d\b{z}^2 +
\frac{2M\b{r}^3}{\b{r}^4 + a^2 \b{z}^2} \left[d\b{t} +
\frac{\b{r}\b{x} + a\b{y}}{\b{r}^2+a^2} d\b{x} + \frac{\b{r} \b{y}
- a\b{x}}{\b{r}^2 + a^2} d\b{y} + \frac{\b{z}}{\b{r}} d\b{z}
\right]^2 \nonumber \\ & = & -d\b{t}^2 + d\b{x}^2 + d\b{y}^2 +
d\b{z}^2 + \frac{2M\b{r}^3}{\b{r}^4 + a^2 \b{z}^2} \left[d\b{t} +
\f{\b{r}}{\b{r}^2 + a^2} \left(\b{x}d\b{x} + \b{y}d\b{y} \right) +
\f{a (\b{y}d\b{x} - \b{x}d\b{y})}{\b{r}^2+a^2} +
\f{\b{z}d\b{z}}{\b{r}} \right]^2\label{eq:KerrKSmetric} \ea
where we rewrote the Kerr metric in the second line in a form that
will be useful in the following. In the $ a \rightarrow 0 $ limit,
we recover the Schwarzschild metric in Kerr-Schild coordinates.
The radial coordinate $ \b{r} $ is related to the fundamental
coordinates $ \b{x}, \b{y}, \b{z} $ by the equation of an oblate
ellipsoid
\be \frac{\b{x}^2 + \b{y}^2} {\b{r}^2 + a^2} +
\frac{\b{z}^2}{\b{r}^2} = 1 \label{eq:ellipsoid} \ee
which is equivalent to a quadratic equation in $ \b{r}^2 $
\be \b{r}^4 - \b{r}^2 (\b{x}^2 + \b{y}^2 + \b{z}^2 - a^2) - a^2
\b{z}^2 = 0 \: ,\label{eq:quadR} \ee
and the horizon is located at $ \b{r} = r_+ = M + \sqrt{M^2 - a^2}
$. Eq. (\ref{eq:ellipsoid}) motivates spheroidal coordinates:

\ba \b{x} & = & \sqrt{\b{r}^2 + a^2} \s{\b{\th}} \c{\b{\ph}}
\nonumber
\\  \b{y} & = & \sqrt{\b{r}^2 + a^2} \s{\b{\th}} \s{\b{\ph}}
\label{eq:BLcoord}
\\  \b{z} & = & \b{r} \c{\b{\th}}. \nonumber \ea
We now explicitly reintroduce the cylindrical coordinates $
(\b{r}_{||}, \b{r}_\perp, \bar{\phi}_{cyl}) $ of the previous
section :
\ba \b{x} & = & \b{r}_{||} \s{\th_\bt} \c{\ph_\bt} + \b{r}_\p
\left(\c{\th_\bt} \c{\ph_\bt} \c{\b{\ph}_{cyl}}
- \s{\ph_\bt} \s{\b{\ph}_{cyl}} \right) \nonumber \\
\b{y} & = & \b{r}_{||} \s{\th_\bt} \s{\ph_\bt} + \b{r}_\p
\left(\c{\th_\bt} \s{\ph_\bt} \c{\b{\ph}_{cyl}} + \c{\ph_\bt}
\s{\b{\ph}_{cyl}} \right)
\label{eq:CarttoPar} \\
\b{z} & = & \b{r}_{||} \c{\th_\bt} - \b{r}_\p \s{\th_\bt}
\c{\b{\ph}_{cyl}} \nonumber \ea
%
%
The angles $ \th_\bt, \ph_\bt $ specify the direction of the
Lorentz boost $ \mathbf{\beta} $ in spherical coordinates based on
$ \b{x}, \b{y}, \b{z} $: $ \mathbf{\bt} = (\bt \s{\th_\bt}
\c{\ph_\bt}, \bt \s{\th_\bt} \s{\ph_\bt}, \bt \c{\th_\bt}) $.
With the coordinate transformation in Eq. (\ref{eq:CarttoPar}) the
Kerr metric becomes
\ba ds^2 & = & -d\b{t}^2 + d\b{r}_{||}^2 + d\b{r}_\p^2 +
\b{r}_\p^2 d\b{\ph}_{cyl}^2 \nonumber \\ & \quad + & \f{2M
\b{r}}{\b{r}^4 + a^2 \left(\b{r}_{||} \c{\th_\bt} - \b{r}_\p
\s{\th_\bt} \c{\b{\ph}_{cyl}}\right)^2} \nonumber \\ &\qquad
\times & \left[d\b{t} + \f{\b{r}}{\b{r}^2 + a^2} \left(\b{x}d\b{x}
+ \b{y}d\b{y} \right) + \f{\b{z}d\b{z}}{\b{r}} + \f{a
\left(\sin\theta_{\bt}\left[\b{r}_\p
d\b{r}_{||}-\b{r}_{||}\:d(\b{r}_\p\s\b{\phi}_{cyl})\right]-\c\theta_\beta
\b{r}_\perp^2 d\b{\phi}_{cyl} \right)}{\b{r}^2+a^2} \right]^2 \: .
\label{eq:KerrKSmetric3} \ea

We now carry out a boost along the selected cylindrical axis.
Unbarred coordinates will denote the boosted observer frame. They
are related to the barred rest-frame coordinates via Eq.
(\ref{eq:LorentzTrans}). After boosting this metric, we will look
at it on an arbitrary $ t = constant $ hypersurface, which we take
as $ t = 0 $ since this choice simplifies the expressions, to
project out the spatial geometry of the hypersurface in which the
apparent horizon lies. This will leave us with $ d\b{t} = -\g \bt
dr_{||} $ and $ \b{r}_{||} = \g r_{||} $. With these changes
substituted into Eq. (\ref{eq:KerrKSmetric3}) we obtain the
spatial part of the boosted Kerr metric on a $ t = 0 $
hypersurface:
\ba ds^2 {\mid}_{t = 0} & = & dr_{||}^2 + dr_\p^2 + r_\p^2
d\ph_{cyl}^2 \nonumber \\ & + & \f{2M \b{r}}{\b{r}^4 + a^2
\left(\g r_{||} \c{\th_\bt} - r_\p
\s{\th_\bt} \c{\ph_{cyl}}\right)^2} \label{eq:KerrKSmetric4} \\
&\qquad \times & \left[-\gamma \beta d r_{||} + \f{\b{r}}{\b{r}^2
+ a^2} \left(\b{x}d\b{x} + \b{y}d\b{y} \right) +
\f{\b{z}d\b{z}}{\b{r}} + \f{a \left(\sin\theta_{\bt} \gamma
\left[r_\p d
r_{||}-r_{||}\:d(r_\p\s\phi_{cyl})\right]-\c\theta_\beta r_\perp^2
d\phi_{cyl} \right)}{\b{r}^2+a^2} \right]^2 \nonumber \ea
%
%
where a few terms involving $ \b{x}\:, \b{y}\:,\b{z}\:, \b{r} $ in
Eqs. (\ref{eq:KerrKSmetric3}), (\ref{eq:KerrKSmetric4}) were left
untouched with the next step in mind. If one wishes, one could
also write all of these terms as functions of $ r_{||}, r_\p $ and
$ \phi_{cyl} $.

Let us remember what we are after; the 2-metric of the boosted
geometry projected out by the condition $ \b{r} = r_+ $. Eq.
(\ref{eq:ellipsoid}) implies
\be
\f{\b{x}d\b{x}+\b{y}d\b{y}}{\b{r}^2+a^2}+\f{\b{z}d\b{z}}{\b{r}^2}
= \left(\f{\b{x}^2+\b{y}^2}{\left(\b{r}^2+a^2\right)^2} +
\f{\b{z}^2}{\b{r}^4} \right) \b{r} d\b{r} \:\: \longrightarrow 0
\quad \mathrm{at} \quad \b{r} =r_+ \: .\label{eq:ellipsoid3} \ee
Thus if $ d\b{r} = 0 $ (e.g. on the horizon, $ \b{r} = r_+ $), the
left hand side of Eq. (\ref{eq:ellipsoid3}) vanishes. This is the
analogue of Eq. (\ref{eq:rdr}) for the Schwarzschild case of
Section \ref{sec:intro}. This simplification reduces the
complexities of Eq. (\ref{eq:KerrKSmetric4}) substantially:
\be ds^2 {\mid}_{t = 0,\: \b{r}=r_+}  =  \left[
\begin{array}{l} dr_{||}^2 + dr_\p^2 + r_\p^2 d\ph_{cyl}^2 + \f{2M r_+}{r_+^4 + a^2
\left(\g r_{||} \c{\th_\bt} - r_\p
\s{\th_\bt} \c{\ph_{cyl}}\right)^2}  \\
\qquad \times  \left[-\gamma \beta d r_{||} + \f{a
\left(\sin\theta_{\bt} \gamma\: \left[r_\p d
r_{||}-r_{||}\:d(r_\p\s\phi_{cyl})\right]-\c\theta_\beta r_\perp^2
d\phi_{cyl} \right)}{r_+^2+a^2} \right]^2
\end{array} \right]_{\b{r}=r_+} \: . \label{eq:KerrKSmetric5}
\ee
However, it is difficult to translate the horizon condition $
\b{r} = r_+ $ into something meaningful in cylindrical
coordinates. Therefore, we must rewrite Eq.
(\ref{eq:KerrKSmetric5}) in spheroidal coordinates to impose the
condition $ \b{r} = r_+ $ to extract the 2-metric of the apparent
horizon. We do this by going back to Eqs. (\ref{eq:CarttoPar}) and
rewriting them as a matrix equation for both boosted and unboosted
coordinates
\be \left(\begin{array}{c} \b{x} \\ \b{y} \\ \b{z} \end{array}
\right) = {\hat{M}} \left( \begin{array}{c} \b{r}_\p \c{\b{\ph}_{cyl}} \\
\b{r}_\p \s{\b{\ph}_{cyl}} \\ \b{r}_{||} \end{array} \right) =
{\hat{M}}
\left( \begin{array}{c} r_\p \c{\ph_{cyl}} \\
r_\p \s{\ph_{cyl}} \\ \g r_{||} \end{array} \right)
\label{eq:CarttoPar2} \ee
where the components of the matrix $ \hat{M} $ can be determined
from Eqs. (\ref{eq:CarttoPar}). The radial coordinate $ \b{r} $ in
Eq. (\ref{eq:BLcoord}) is related to the Cartesian and cylindrical
coordinates via $ \b{x}^2 + \b{y}^2 + \b{z}^2 = \b{r}_\p^2 +
\b{r}_{||}^2 = \b{r}^2 + a^2 \s^2 \b{\th} $. Since $ \b{r}_{||} =
\g r_{||} $ on the $ t= 0 $ hypersurface, we also have $ r_\p^2 +
\g^2 r_{||}^2 = \b{r}^2 + a^2 \s^2 \b{\th} $ (cf. \cite{Po}).
Setting Eq. (\ref{eq:CarttoPar2}) equal to Eq. (\ref{eq:BLcoord})
and multiplying by $ \hat{M}^{-1} $, we get
\[ \left( \begin{array}{c} r_\p \c{\ph_{cyl}} \\ r_\p
\s{\ph_{cyl}} \\ \g r_{||} \end{array} \right) = {\hat{M}^{-1}}
\left(\begin{array}{c} \sqrt{\b{r}^2 + a^2} \s{\b{\th}}
\c{\b{\ph}} \\ \sqrt{\b{r}^2 + a^2} \s{\b{\th}} \s{\b{\ph}}
\\ \b{r} \c{\b{\th}} \end{array} \right). \]
Expanding this we obtain
\ba r_\p \c{\ph_{cyl}} & = & \c{\th_\bt} \c{\ph_\bt}\sqrt{\b{r}^2
+ a^2} \s{\b{\th}} \c{\b{\ph}} + \c{\th_\bt} \s{\ph_\bt}
\sqrt{\b{r}^2 + a^2} \s{\b{\th}} \s{\b{\ph}} - \s{\th_\bt} \;
\b{r} \c{\b{\th}} \nonumber \\ r_\p \s{\ph_{cyl}} & = & -
\s{\ph_\bt} \sqrt{\b{r}^2 + a^2}  \s{\b{\th}} \c{\b{\ph}} +
\c{\ph_\bt}
\sqrt{\b{r}^2 + a^2} \s{\b{\th}} \s{\b{\ph}}  \label{eq:cyltoSph} \\
\g r_{||} & = & \s{\th_\bt} \c{\ph_\bt}\sqrt{\b{r}^2 + a^2}
\s{\b{\th}} \c{\b{\ph}} + \s{\th_\bt} \s{\ph_\bt} \sqrt{\b{r}^2 +
a^2} \s{\b{\th}} \s{\b{\ph}} + \c{\th_\bt}\; \b{r} \c{\b{\th}} \:
. \nonumber  \ea
\par In the limit $ \th_\bt = \ph_\bt = 0 $, the equations above
reduce to Eq. (\ref{eq:BLcoord}) with the cylindrical coordinates
replacing $ (\b{x}, \b{y}, \b{z}) $. This is the case of boosting
along the z-axis, and we briefly treat that here before
proceeding.
For boost along the z-axis, with $ \b{r} = r_+ $ (i.e. on the
horizon) we have
\be ds^2{\mid}_{t=0,\:\b{r} = r_+} = \left[dx^2 + dy^2 +dz^2 +
\f{r_+^2 \: (r_+^2 + a^2)}{r_+^4 + \gamma^2 a^2 z^2} \left[-\gamma
\beta dz + \f{a(y dx - x dy)}{r_+^2 + a^2} \right]^2
\right]_{\b{r} = r_+}\: . \label{eq:KerrKS_Zboost2} \ee
Because of the boost in the z-direction, only the terms involving
z ($ -\gamma\beta dz $ in the numerator and $ \gamma^2 a^2 z^2 $
in the denominator) differ from the unboosted case. In Eq.
(\ref{eq:KerrKS_Zboost2}) we still have to evaluate some of the
terms at $ \b{r} = r_+ $. Using spheroidal coordinates
\ba     x & = & \sqrt{r_+^2 + a^2} \sin\b{\theta} \cos\b{\phi} \nonumber \\
   y & = & \sqrt{r_+^2 + a^2} \sin\b{\theta} \sin\b{\phi}\\
   \gamma z & = & r_+ \cos\b{\theta} \nonumber \ea
we get
\ba ds^2{\mid}_{t=0, \:\b{r} = r_+} & = & (r_+^2+a^2)
(\cos^2\b{\theta} d\b{\theta}^2 +
\sin^2 \b{\theta} d\b{\phi}^2) + \f{r_+^2}{\gamma^2} \sin^2\b{\theta} d\b{\theta}^2 \nonumber \\
       \quad \quad & + & \f{r_+^2 + a^2}{r_+^2 + a^2 \cos^2 \b{\theta}}
       \left[ \beta r_+ \sin\b{\theta} d\b{\theta} - a \sin^2\b{\theta} d\b{\phi}
       \right]^2\: .
      \label{eq:KerrKS_Zboost3} \ea
With further simplifications, this becomes
\be ds^2{\mid}_{t=0, \:\b{r} = r_+} = (r_+^2 + a^2
\cos^2\b{\theta}) d\b{\theta}^2 +\f{\sin^2\b{\theta}}{r_+^2 + a^2
\cos^2\bar{\theta}} \left[-\beta a r_+ \s\b{\theta}d\b{\theta} +
(r_+^2 + a^2) d\b{\phi}\right]^2 \: . \label{eq:KerrKS_Zboost4}
\ee
In the $ a\rightarrow 0 $ limit,  Eq. (\ref{eq:KerrKS_Zboost4})
gives precisely the expression we obtained for the boosted
Schwarzschild metric. Let us now look at the 2-metric $ g_{AB} (A,
B = \theta, \phi) $ for the apparent horizon component by
component.
\ba g_{\b{\theta} \b{\theta}} & = & \left(r_+^2 + a^2
\cos^2\b{\theta}\right) +
\left(\f{\gamma\beta a r_+ \sin^2\b{\theta}}{\sqrt{r_+^2 + a^2 \cos^2\b{\theta}}}\right)^2 \: ,\nonumber \\
g_{\b{\phi} \b{\phi}} & = & \left(\f{(r_+^2+a^2) \sin\b{\theta}}{\sqrt{r_+^2 + a^2 \cos^2\b{\theta}}}\right)^2 \: , \nonumber \\
g_{\b{\theta} \b{\phi}} & = & -\left(\f{(r_+^2+a^2)
\sin\b{\theta}}{\sqrt{r_+^2 + a^2 \cos^2\b{\theta}}}\right)
\left(\f{\gamma\beta a r_+ \sin^2\b{\theta}}{\sqrt{r_+^2 + a^2
\cos^2\b{\theta}}}\right)\: . \label{eq:2_metricZ} \ea
For any $ 2 \times 2 $ matrix of the form
\be H_{AB} = \left(\begin{array}{cc} A^2+B^2 & BC \\ BC & C^2
\end{array}\right) \label{eq:matrix} \ee
the determinant is $ det H_{AB} = A^2 C^2 $. The 2-dimensional
metric is of this form, so
\be \sqrt{det\left(g_{AB}\right)} = (r_+^2 + a^2) \sin\b{\theta}
\ee
Since
\be \mathrm{Area} =\int \sqrt{det\left(g_{AB}\right)} d\b{\theta}
d\b{\phi} \label{eq:Area} \ee
we obtain an area of $ 4\pi \left(r_+^2 + a^2 \right) $ as
expected, identical to the unboosted horizon area.

Going back to our boost in an arbitrary direction, we rewrite the
3-metric in Eq. (\ref{eq:KerrKSmetric5}) using the spheroidal
coordinates of Eq. (\ref{eq:cyltoSph}). After some algebra using a
well known algebraic relation for the Kerr spacetime ($ 2M r_+ =
r_+^2 + a^2 $) to simplify, and setting $\b{r} = r_+$ in most
places, we end up with a result surprisingly similar to Eq.
(\ref{eq:KerrKS_Zboost4})
\be ds^2 {\mid}_{t = 0, \b{r} = r_+}  =  \left( r_+^2 + a^2 cos^2
\b{\theta} \right) d \b{\theta}^2 + \f{\s^2 \b{\th}}{r_+^2 + a^2
\c^2 \bar{\th}} \left[ a \g \bt dr_{||} + \left(r_+^2 + a^2
\right) d \b{\ph} \right]^2 \label{eq:KerrKSmetric6} \ee
%
Using the last one of Eqs. (\ref{eq:cyltoSph}), we now expand the
terms containing $ dr_{||} $ and obtain the components of the
2-metric for the apparent horizon:
\ba g_{\t{\th} \t{\th}} & = & \left(r_+^2 + a^2 \c^2 \t{\th}
\right) \nonumber \\  & \qquad & + \f{\bt^2 a^2 \s^2
\t{\th}}{r_+^2 + a^2 \c^2 \t{\th}} \left( \sqrt{ r_+^2 + a^2}\s
\th_\bt \c \t{\th} \c (\t{\ph} - \ph_\bt ) - r_+ \c \th_\bt \s
\t{\th} \right)^2 \: , \label{eq:g_thth} \ea
\be g_{\t{\ph} \t{\ph}}  =  \f{\left( r_+^2 + a^2 \right) \s^2
\t{\th}}{r_+^2 + a^2 \c^2 \t{\th}} \left( \sqrt{r_+^2 + a^2} - \bt
a \s \th_\bt \s \t{\th} \s (\t{\ph} - \ph_\bt )\right)^2 \: ,
\label{eq:g_phph} \ee
\be g_{\t{\th} \t{\ph}} = \f{\bt a \sqrt{r_+^2 + a^2} \s^2
\t{\th}}{r_+^2 + a^2 \c^2 \t{\th}} \left[
\begin{array}{l} \left( \sqrt{ r_+^2 + a^2}\s
\th_\bt \c \t{\th} \c (\t{\ph} - \ph_\bt ) - r_+ \c \th_\bt \s
\t{\th} \right) \\ \:\:\times \left( \sqrt{r_+^2 + a^2} - \bt a \s
\th_\bt \s \t{\th} \s (\t{\ph} - \ph_\bt )\right)
\end{array} \right] \: . \label{eq:metricElements}\ee
%
%
The $ \beta \rightarrow 0 $ limit of equations (\ref{eq:g_thth})
through (\ref{eq:g_phph}) yields the standard 2-metric of the Kerr
spacetime given in Boyer-Lindquist coordinates. The $ a
\rightarrow 0 $ limit gives the standard Schwarzschild (spherical)
2-metric. The $ \theta_\beta = 0 $ limit yields the metric of Eq.
(\ref{eq:2_metricZ}). Eqs.
(\ref{eq:g_thth})-(\ref{eq:metricElements}) show that $ g_{AB} $
is again of the form of Eq. (\ref{eq:matrix}). Hence the square
root of the determinant of the metric is
\be \sqrt{\det \left(g_{AB}\right)} = \left(r_+^2 + a^2\right)
\s{\b{\th}} - \bt a \sqrt{r_+^2 + a^2} \s{\th_\bt} \s^2 \b{\th}
\s{(\b{\ph}-\ph_\bt)}\: . \label{eq:sqrtDet} \ee
The first term above is the familiar contribution from the
unboosted Kerr metric. To determine the area, we integrate the
square root of the determinant of the 2-metric over the angular
variables of the spheroidal coordinate system.
\begin{eqnarray}
\mathrm{Area} & = & \int \sqrt{\det \left(g_{AB}\right)}d\b{\theta}d\b{\phi} \nonumber \\
    & = & \int_0^{2 \pi} d\b{\ph} \int_0^{\pi} d\b{\th}
    \left[\left(r_+^2 + a^2\right) \s{\b{\th}} - \bt a
\sqrt{r_+^2 + a^2} \s{\th_\bt} \s^2 \b{\th}
\s{(\b{\ph}-\ph_\bt)}\right]
    \nonumber \\
    & = & 4 \pi \left( r_+^2 + a^2 \right) \: .
\end{eqnarray}
Above, the second term disappears because of the $ \b{\ph} $
integral. Our calculation shows that the area of the apparent
horizon of a Kerr black hole remains invariant under arbitrary
Lorentz boosts, as expected.

\section{Conclusions}
\label{sec:Conclusions} Our goal was to show that the area of the
apparent horizons of Kerr black holes remain invariant under a
particular redefinition of the $ t = constant $ hypersurface
(Lorentz boosts in arbitrary directions on the Kerr-Schild form).
We introduced boost-parallel cylindrical coordinates. In this
form, it is almost trivial to boost the spacetime metric. Once
boosted, we looked at a $ t = constant $ hypersurface to determine
the three dimensional spatial portion of the boosted metric, and
projected down to the 2-metric of the apparent horizon. We gave
examples and validation of using these results based on
computational initial data, to obtain binding energy results for
nonspinning black holes boosted together or apart. The binding
energy curves with different $ \beta$ values closely overlap and
agree in the Newtonian limit; the slight deviation of the binding
energy from the $ 1/ r $ form can plausibly be explained by
nonlinear corrections to the physical separation corresponding to
a given coordinate separation. This is a subject of a
post-Newtonian study in progress.

For the general direction boosted Kerr case the 2-metric of the
apparent horizon has non-zero off-diagonal terms; however, when
integrated over the angular variables, these contribute zero to
the area, leaving us with the same result as the undisturbed Kerr
case, namely $ Area = 4 \pi \left( r_+^2 + a^2 \right) $.

\par We performed all of our calculations
in one particular type of slicing of the spacetime, i.e. $ t =
constant $. There are infinitely many slicings of static and
stationary spacetimes, including ones with no apparent horizon at
all \cite{WaII}. But if the $ t = constant $  space contains an
apparent horizon then the horizon area has the standard value.
Black holes are very special objects indeed.

\begin{acknowledgements}
This work was supported by NSF grant PHY-0354842 and NASA grant
NNG 04GL37G. Sarp Akcay would like to thank Cihan Akcay for his
assistance with MATlab.
\end{acknowledgements}


\end{document}